\documentclass[twocolumn,showpacs,amsmath,amssymb,amsfonts,a4paper,aps,prd]{revtex4}
\usepackage{graphicx,color}

\newcommand{\f}[2]{\frac{#1}{#2}}
\def\be{\begin{equation}}
\def\ee{\end{equation}}
\def\bea{\begin{eqnarray}}
\def\eea{\end{eqnarray}}

\begin{document}
\title{Weyl-Cartan-Weitzenb\"{o}ck gravity through Lagrange multiplier}
\author{Zahra Haghani$^1$}
\email{z_haghani@sbu.ac.ir}
\author{Tiberiu Harko$^2$}
\email{t.harko@ucl.ac.uk}
\author{Hamid Reza Sepangi$^1$}
\email{hr-sepangi@sbu.ac.ir}
\author{Shahab Shahidi$^1$}
\email{s_shahidi@sbu.ac.ir}
\affiliation{$^1$Department of Physics, Shahid Beheshti University, G. C.,
Evin,Tehran 19839, Iran}
 \affiliation{$^2$Department of Mathematics, University College London, Gower Street, London, WC1E 6BT, United Kingdom}
\begin{abstract}
We consider an extension of the Weyl-Cartan-Weitzenb\"{o}ck (WCW) and teleparallel gravity, in which the Weitzenb\"{o}ck condition of the exact cancellation of curvature and torsion in a Weyl-Cartan geometry is inserted into the gravitational action via a  Lagrange multiplier.
  In the standard metric formulation of the WCW model, the flatness of the  space-time is removed by imposing the Weitzenb\"{o}ck condition in the Weyl-Cartan geometry, where the dynamical variables are the space-time metric, the Weyl vector and the torsion tensor, respectively. However, once the Weitzenb\"{o}ck condition is imposed on the Weyl-Cartan space-time,  the metric is not dynamical, and the gravitational  dynamics and evolution is completely determined by the torsion tensor. We show how to resolve this difficulty, and generalize the WCW model, by imposing the Weitzenb\"{o}ck condition on the action of the gravitational field through a Lagrange multiplier. The gravitational field equations are obtained from the variational principle, and they explicitly depend on the Lagrange multiplier. As a particular model we  consider the case of the Riemann-Cartan space-times with zero non-metricity, which mimics the teleparallel theory of gravity. The Newtonian limit of the model is investigated, and a generalized Poisson equation is
obtained, with the weak field gravitational potential explicitly depending on the Lagrange multiplier and on the Weyl vector. The cosmological implications of the theory are also studied, and three classes of exact cosmological models are considered.
\end{abstract}
\maketitle

                              \section{Introduction}\label{intro}
General relativity (GR) is considered to be the most successful theory of gravity ever proposed. Its classic predictions on the perihelion advance of Mercury, on the deflection of light by the Sun, gravitational redshift, or radar echo delay have been confirmed at an unprecedented level of observational accuracy. Moreover,  predictions such as the orbital decay of the Hulse-Taylor binary pulsar, due to gravitational - wave damping, have also fully confirmed the observationally  weak-field validity of the theory. The detection of the gravitational waves will allow the testing of the predictions of GR in the strong gravitational field limit, such as, for example,  the final stage of binary black hole coalescence (for a recent review on the experimental tests of GR see \cite{Will}).

Despite these important achievements, recent observations of supernovae \cite{Riess} and of the Cosmic Microwave Background radiation \cite{1} have suggested that  on cosmological scales GR may not be the ultimate theory to describe the Universe. If GR is correct, in order to explain the accelerating
expansion of the Universe, we require that the Universe is filled with some component of unknown nature, called dark energy, having some unusual physical properties. To find an alternative to dark energy to explain cosmological observations, in the past decade many modified theories of gravity, which deviate from the
standard GR on cosmological scales have been proposed (see \cite{Clifton} for a recent review on  modified gravity and cosmology). On the other hand, because of its prediction of space-time singularities in the Big Bang and  inside black holes  GR could be considered  as an incomplete physical model. In order to solve the singularity problem  it is generally believed that a consistent extension of GR into the quantum domain is needed.

Since GR is essentially a geometric theory, formulated in the Riemann space, looking for more general geometric structures adapted for the description of the gravitational field may be one of the most promising ways for the explanation of the behavior at large cosmological scales  of the matter in the  Universe, whose structure and dynamics  may be described by more general geometries than the Riemannian one, valid at the Solar System level.

The first attempt to create a more general geometry is due to Weyl \cite{Weyl}, who proposed a geometrized unification of gravitation and electromagnetism. Weyl abandoned the metric-compatible Levi-Civita connection as a fundamental concept, since it allowed the distant comparison of lengths.  Substituting the metric field by the class of all conformally equivalent metrics, Weyl  introduced a connection that would not carry any information about the length of a vector on parallel transport. Instead the latter task was assigned to an extra connection, a so-called length connection that would, in turn, not carry any information about the direction of a vector on parallel transport, but that would only fix, or gauge, the conformal factor. Weyl identified the length connection with the electromagnetic potential. A generalization of Weyl’s theory was introduced by Dirac \cite{Dirac}, who proposed the existence of two metrics, one unmeasurable metric $ds_E$, affected by transformations in
the standards of length, and a second measurable one, the conformally invariant atomic metric $ds_A$.

In the development of the generalized geometric theories of gravity a very different evolution took place  due to the work of Cartan \cite{Cartan}, who proposed an extension of general relativity, which is known today as the Einstein-Cartan theory \cite{Hehl1}. The new geometric element of the theory, the torsion field, is usually associated from a physical point of view to a spin density \cite{Hehl1}. The Weyl geometry can be immediately generalized to include the torsion. This geomety is called the Weyl-Cartan space-time, and it was extensively studied from both mathematical and physical points
of view \cite{WC}. To build up an action
integral from which one can obtain a gauge covariant (in the Weyl sense) general relativistic massive electrodynamics, torsion was included in the geometric framework of the Weyl-Dirac theory in \cite{Isr} .
For a recent review of the geometric properties and of the physical applications of the Riemann-Cartan and Weyl-Cartan space-times see \cite{Rev}.

A third independent mathematical development took place in  the work of Weitzenb\"{o}ck \cite{Weitz}, who introduced the so-called Weitzenb\"{o}ck spaces. A Weitzenb\"{o}ck manifold has the properties $\nabla _{\mu }g_{\sigma \lambda }= 0$, $T^{\mu }_{\sigma \lambda }\neq 0$, and $R^{\mu }_{\nu \sigma \lambda }=0$, where $g_{\sigma \lambda }$, $T^{\mu }_{\sigma \lambda }$ and $R^{\mu }_{\nu \sigma \lambda }$ are the metric, the torsion, and the curvature tensors of
the manifold, respectively. When $T^{\mu }_{\sigma \lambda }= 0$, the manifold is reduced to a Euclidean manifold. The torsion tensor possesses different values  on different parts of the Weitzenb\"{o}ck manifold. Therefore, since their Riemann curvature tensor is zero, Weitzenb\"{o}ck spaces possess the property of distant parallelism, also known as absolute, or teleparallelism. Weitzenb\"{o}ck type geometries  were first used in physics by Einstein, who proposed a unified teleparallel theory of gravity and electromagnetism  \cite{Ein}. The basic idea of the teleparallel approach  is to substitute, as a basic physical variable, the metric $g_{\mu \nu}$ of the space-time by a set of tetrad vectors $e^i_{\mu }$.
 In this approach the torsion, generated by the tetrad fields, can be used to describe general relativity entirely,
with the  curvature eliminated in favor of torsion. This is the so-called teleparallel equivalent of General Relativity (TEGR),
which was introduced in \cite{11}, and is also known as the $f(T)$ gravity model. Therefore, in teleparallel, or f(T) gravity,   torsion
exactly compensates curvature, and the space-time becomes flat. Unlike in $f(R)$ gravity, which in the metric approach is a fourth order theory, in the $f(T)$ gravity models the field equations are of second order. $f(T)$ gravity models have been
extensively applied to cosmology, and in particular to explain the late-time accelerating expansion of the Universe, without the need
of dark energy \cite{13}.

An extension of the teleparallel gravity models, called WCW gravity, was introduced recently in \cite{WCW}. In this approach, the Weitzenb\"{o}ock condition of the vanishing of the sum of the curvature and torsion scalar is imposed in a background  Weyl-Cartan type space-time. In contrast to the
standard teleparallel theories, the model is formulated in a four-dimensional curved space-time, and not in a flat Euclidian geometry. The
properties of the gravitational field are  described by the torsion tensor and the Weyl vector fields, defined in a four-dimensional curved space-time manifold. In the gravitational action a kinetic term for the torsion is also included. The field equations of the
model, obtained from a Hilbert-Einstein type variational principle, allow a complete
description of the gravitational field in terms of two vector fields, the Weyl vector and torsion, respectively, defined in a curved background.
The Newtonian limit of the model was also considered, and it was shown that in the weak gravitational field approximation the standard Poisson
equation can be recovered. For a particular choice of the free parameters, in which the torsion vector is proportional to the
Weyl vector, the cosmological applications of the model were  investigated. A large variety of dynamical evolutions can be obtained in the WCW gravity model, ranging from inflationary/accelerated expansions to non-inflationary behaviors. The nature of the cosmological evolution is determined by the numerical values of the parameters of the cosmological model. In particular  a de Sitter type late time evolution can be naturally obtained from the field equations of the model. Therefore the WCW gravity model leads to the possibility of a purely geometrical description
of dark energy  where the late time acceleration of the Universe is determined by the intrinsic
nature of the space-time.

Recently, the use of Lagrange multipliers in the formulation of dynamical gravity models has attracted considerable attention.  The method of Lagrange multipliers  is a strategy for finding the local maxima and minima of a function subject to equality nonholonomic constraints, which are capable of reducing the dynamics \cite{20}. The extension of $f(R)$ gravity models via the addition of
a Lagrange multiplier constraint has been proposed in \cite{fR}. This model can be considered as a new version of $f(R)$ modified gravity since dynamics, and the cosmological solutions, are different from the standard version of $f(R)$ gravity without such constraint. Cosmological models with Lagrange multipliers  have been considered from different points of view in \cite{LM}.

It is the purpose of the present paper to investigate a class of generalized WCW type gravity models, in which the Weitzenb\"{o}ck condition of the exact compensation of torsion and curvature is introduced into the gravitational action via a Lagrange multiplier approach. We start our analysis by considering the general action for a gravitational field in a Weyl-Cartan space-time, and we explicitly introduce the  Weitzenb\"{o}ck condition into the action via a Lagrange multiplier. By taking the Weyl vector as being identically zero, we obtain the field equations of this gravity model in a Riemann-Cartan space time, with the  Weitzenb\"{o}ck condition being described by a proportionality relation between the scalar curvature and torsion scalar, included in the gravitational action via a Lagrange multiplier method which mimics the teleparallel gravity. The weak field limit of the general theory is also investigated, and a
generalized Poisson equation, explicitly depending on the Lagrange multiplier and the Weyl vector is obtained.

The cosmological implications of the model are investigated for three classes of models. The  solutions obtained describe both accelerating and decelerating expansionary phases of the Universe, and they may prove useful for modeling the early and late phases of cosmological evolution.

The  paper is organized as follows. The gravitational action of the WCW theory with Lagrange multiplier is introduced in Section~\ref{model}. The gravitational field equations are derived in Section~\ref{feq}. Some particular cases are also considered in detail. The field equations for the case of the zero Weyl vector are presented in Section~\ref{w}.  The weak field limit of the theory is investigated in Section~\ref{weak}, and the generalized Poisson equation is obtained. The cosmological implications of the theory are investigated in Section~\ref{cos}, and several cosmological models are presented. We discuss and conclude our results in Section~\ref{conclu}.  Some aspects of the Weyl invariance of the theory are considered in the Appendix.

                              \section{WCW gravity model with Lagrange multiplier}\label{model}
In this Section we formulate the action of the gravitational field in the WCW gravity with a Lagrange multiplier. A Weyl - Cartan space $CW_4$ is a four-dimensional connected, oriented, and differentiable manifold, having
a metric with a Lorentzian signature chosen as $(-+++)$, curvature, torsion, and a connection  which can be determined from the Weyl nonmetricity condition. Hence the Weyl-Cartan geometry has the properties that the connection is no longer symmetric, and the metric compatibility condition does not hold. The Weyl non-metricity condition is defined as
\begin{align}\label{eq1}
\nabla_\lambda g_{\mu\nu}=-2w_\lambda g_{\mu\nu},
\end{align}
where $w_\mu$ is the Weyl vector. Expanding the covariant derivative, we obtain the connection in the Weyl-Cartan geometry
\begin{align}\label{eq2}
\Gamma^\lambda_{~\mu\nu}=\left\{\begin{matrix}\lambda\\ \mu~\nu\end{matrix}\right\}+C^\lambda_{\mu\nu}+g_{\mu\nu}w^\lambda-\delta^\lambda_\mu w_\nu-\delta^\lambda_\nu w_\mu,
\end{align}
where the first term in the LHS is the Christoffel symbol constructed out of the metric and the contorsion tensor $C^\lambda_{~\mu\nu}$ is defined  as
\begin{align}\label{eq3}
C^\lambda_{~\mu\nu}=T^\lambda_{~\mu\nu}-g^{\lambda\beta}g_{\sigma\mu}T^\sigma_{~\beta\nu}-g^{\lambda\beta}g_{\sigma\nu}T^\sigma_{~\beta\mu},
\end{align}
with torsion tensor $T^\lambda_{~\mu\nu}$ given by
\begin{align}\label{eq4}
T^\lambda_{~\mu\nu}=\f{1}{2}\left(\Gamma^\lambda_{~\mu\nu}-\Gamma^\lambda_{~\nu\mu}\right).
\end{align}
One can then obtain the curvature tensor of the Weyl-Cartan space-time  as
\begin{align}\label{eq5}
K^\lambda_{~\mu\nu\sigma}=\Gamma^\lambda_{~\mu\sigma,\nu}-\Gamma^\lambda_{~\mu\nu,\sigma}+\Gamma^\alpha_{~\mu\sigma}\Gamma^\lambda_{~\alpha\nu}-\Gamma^\alpha_{~\mu\nu}\Gamma^\lambda_{~\alpha\sigma}.
\end{align}

Using  equation (\ref{eq2}) and contracting the curvature tensor with the metric, we obtain the curvature scalar
\begin{align}\label{eq6}
K=K^{\mu\nu}_{~~\mu\nu}&=R+6\nabla_\nu w^\nu-4\nabla_\nu T^\nu-6w_\nu w^\nu+8 w_\nu T^\nu\nonumber\\
&+T^{\mu\alpha\nu}T_{\mu\alpha\nu}+2T^{\mu\alpha\nu}T_{\nu\alpha\mu}-4T_\mu T^\mu,
\end{align}
where $R$ is the curvature scalar constructed from the Christoffel symbols and we have defined $T_\beta=T^\alpha_{~\beta\alpha}$. Also all covariant derivatives are with respect to the Riemannian connection described by the Christoffel symbols constructed out of the metric $g_{\mu \nu}$. We also introduce two tensor fields $W_{\mu\nu}$ and $T_{\mu\nu}$, constructed from the Weyl vector and the torsion vector, respectively
\begin{align}\label{eq8}
W_{\mu\nu}&=\nabla_\nu w_\mu-\nabla_\mu w_\nu,\\
T_{\mu\nu}&=\nabla_\nu T_\mu-\nabla_\mu T_\nu,
\end{align}
where $T=T_\mu T^\mu$.

The most general action for a gravitational theory in the Weyl-Cartan space-time can then be formulated as
\begin{align}\label{eq7}
S=\int d^4x\sqrt{-g}\bigg(\f{1}{\kappa^2}K&-\f{1}{4}W_{\mu\nu}W^{\mu\nu}\nonumber\\&+\hat\beta\nabla_\mu T\nabla^\mu T+\hat\alpha T_{\mu\nu}T^{\mu\nu}+L_m\bigg),
\end{align}
where $L_m$ is the matter Lagrangian which depends only on the matter fields and the metric, and is independent on the torsion tensor and the Weyl vector. We have also added a kinetic term for the Weyl vector and two possible kinetic terms for the torsion tensor. In equation (\ref{eq7}), $\hat{\alpha}$ and $\hat{\beta }$ are arbitrary numerical constants, and $\kappa ^2=16\pi G$.
Substituting  definition of the curvature scalar from equation \eqref{eq6}, the  action for the gravitational field becomes
\begin{align}\label{eq9}
S=&\f{1}{\kappa^2}\int d^4x\sqrt{-g}\bigg(R+T^{\mu\alpha\nu}T_{\mu\alpha\nu}+2T^{\mu\alpha\nu}T_{\nu\alpha\mu}\nonumber\\
&-4T_\mu T^\mu-6w_\nu w^\nu+8 w_\nu T^\nu-\f{\kappa^2}{4}W_{\mu\nu}W^{\mu\nu}\nonumber\\
&+\beta\nabla_\mu T\nabla^\mu T+\alpha T_{\mu\nu}T^{\mu\nu}+\kappa^2L_m\bigg),
\end{align}
where we have defined $\alpha=\kappa^2\hat\alpha$ and $\beta=\kappa^2\beta$.

The  Weitzenb\"{o}ck condition
\begin{align}\label{eq10}
\mathcal{W}\equiv R+T^{\mu\alpha\nu}T_{\mu\alpha\nu}+2T^{\mu\alpha\nu}T_{\nu\alpha\mu}-4T_\mu T^\mu=0,
\end{align}
requires that the sum of the scalar curvature and torsion be zero.
In order to impose this condition on the gravitational field equations of the theory, we  add it to the action by  using a Lagrange multiplier $\lambda$. The gravitational action then becomes
\begin{align}\label{eq11}
S&=\f{1}{\kappa^2}\int d^4x\sqrt{-g}\bigg[-\f{\kappa^2}{4}W_{\mu\nu}W^{\mu\nu}-6w_\nu w^\nu+8 w_\nu T^\nu\nonumber\\
&+(1+\lambda)\big(R+T^{\mu\alpha\nu}T_{\mu\alpha\nu}+2T^{\mu\alpha\nu}T_{\nu\alpha\mu}-4T_\mu T^\mu\big)\nonumber\\
&+\beta\nabla_\mu T\nabla^\mu T+\alpha T_{\mu\nu}T^{\mu\nu}+\kappa^2L_m\bigg].
\end{align}
 We note that for $\lambda=-1$ one gets the original WCW action \cite{WCW}.

\section{The gravitational  field equations of the WCW gravity model with a Lagrange multiplier}\label{feq}

Let us now derive the field equations of the WCW gravity with the Lagrange multiplier. By considering the irreducible decomposition of the torsion tensor  and  imposing a condition on the terms of the decomposition, we obtain an explicit representation of the Weitzenb\"{o}ck condition. The field equations of a simplified model in which the constant $\alpha =0$ are also obtained explicitly.

\subsection{The gravitational field equations and the effective energy-momentum tensor}

Variation of the action \eqref{eq11} with respect to the Weyl vector and the torsion tensor results in the equations of motion
\begin{align}\label{eq12}
-\f{\kappa^2}{2}\nabla_\nu W^{\nu\mu}-6w^\mu+4T^\mu=0,
\end{align}
and
\begin{align}\label{eq13}
4w^{[\rho}&\delta^{\beta]}_\mu+2\alpha\delta^{[\beta}_\mu\nabla_\alpha T^{\rho]\alpha}-2\beta T^{[\rho}\delta^{\beta]}_\mu\Box T\nonumber\\
&+(1+\lambda)\big(T_\mu^{~\rho\beta}+T^{\beta\rho}_{~~\mu}+T^{\rho~\beta}_{~\mu}-4T^{[\rho}\delta^{\beta]}_\mu\big)=0,
\end{align}
respectively. Variation of the action with respect to the Lagrange multiplier $\lambda$ gives the Weitzenb\"{o}ck condition \eqref{eq10}. Now, varying the action with  respect to the metric and using the condition \eqref{eq10}, we obtain the dynamical equation for the metric as
\begin{widetext}
\begin{align}\label{eq14}
(1+\lambda)R_{\mu\nu}&=\f{\kappa^2}{2}T^m_{\mu\nu}+\nabla_\mu\nabla_\nu\lambda-g_{\mu\nu}\Box\lambda-(1+\lambda)\bigg(2T^{\alpha\beta}_{~~~\nu}T_{\alpha\beta\mu}-T_\mu^{~\alpha\beta}T_{\nu\alpha\beta}+2T_{\beta\alpha(\mu}T^{\alpha\beta}_{~~~\nu)}-4T_\mu T_\nu\bigg)\nonumber\\
&+\f{\kappa^2}{2}\left(W_{\mu\alpha}W_\nu^{~\alpha}-\f{1}{4}W_{\alpha\beta}W^{\alpha\beta}g_{\mu\nu}\right)-2\alpha\left(T_{\mu\alpha}T_\nu^{~\alpha}-\f{1}{4}T_{\alpha\beta}T^{\alpha\beta}g_{\mu\nu}\right)+6\left(w_\mu w_\nu-\f{1}{2}w^\alpha w_\alpha g_{\mu\nu}\right)\nonumber\\
&-\beta\left(\nabla_\mu T\nabla_\nu T-\f{1}{2}g_{\mu\nu}\nabla_\alpha T\nabla^\alpha T-2 T_\mu T_\nu\Box T\right)-8\left(T_{(\mu}w_{\nu)}-\f{1}{2}w^\alpha T_\alpha g_{\mu\nu}\right),
\end{align}
where $T^m_{\mu\nu}$ is the energy-momentum of the ordinary-matter. The generalized Einstein field equation \eqref{eq14} can be written as
\begin{align}\label{eq14-1}
G_{\mu\nu}=T^{eff}_{\mu\nu},
\end{align}
where we have defined the effective energy-momentum tensor as
\begin{align}\label{eq14-2}
T^{eff}_{\mu\nu}&=(1+\lambda)^{-1}\bigg[ \f{\kappa^2}{2}T^m_{\mu\nu}+\nabla_\mu\nabla_\nu\lambda-g_{\mu\nu}\Box\lambda-(1+\lambda)\bigg(2T^{\alpha\beta}_{~~~\nu}T_{\alpha\beta\mu}-T_\mu^{~\alpha\beta}T_{\nu\alpha\beta}+2T_{\beta\alpha(\mu}T^{\alpha\beta}_{~~~\nu)}-4T_\mu T_\nu\bigg)\nonumber\\
&+\f{\kappa^2}{2}\left(W_{\mu\alpha}W_\nu^{~\alpha}-\f{1}{4}W_{\alpha\beta}W^{\alpha\beta}g_{\mu\nu}\right)-2\alpha\left(T_{\mu\alpha}T_\nu^{~\alpha}-\f{1}{4}T_{\alpha\beta}T^{\alpha\beta}g_{\mu\nu}\right)+6\left(w_\mu w_\nu-\f{1}{2}w^\alpha w_\alpha g_{\mu\nu}\right)\nonumber\\
&-\beta\left(\nabla_\mu T\nabla_\nu T-\f{1}{2}g_{\mu\nu}\nabla_\alpha T\nabla^\alpha T-2 T_\mu T_\nu\Box T\right)-8\left(T_{(\mu}w_{\nu)}-\f{1}{2}w^\alpha T_\alpha g_{\mu\nu}\right)\nonumber\\&+\f{1}{2}(1+\lambda)\left(T^{\gamma\beta\alpha}T_{\gamma\beta\alpha}+2T^{\gamma\beta\alpha}T_{\alpha\beta\gamma}-4T^\alpha T_\alpha\right)g_{\mu\nu}\bigg],
\end{align}
and use has been made of equation \eqref{eq10}.
\end{widetext}

\subsection{The decomposition of the torsion tensor}

The torsion tensor can be decomposed irreducibly into
\begin{align}\label{eq15}
T_{\mu\nu\rho}=\f{2}{3}(t_{\mu\nu\rho}-t_{\mu\rho\nu})+\f{1}{3}(Q_\nu g_{\mu\rho}-Q_\rho g_{\mu\nu})+\epsilon_{\mu\nu\rho\sigma}S^\sigma,
\end{align}
where $Q_\nu$ and $S^\rho$ are two vectors, and the tensor $t_{\mu\nu\rho}$ is symmetric under the change of the first two indices, and satisfies the following conditions
\begin{align}\label{eq16}
t_{\mu\nu\rho}+t_{\nu\rho\mu}+t_{\rho\mu\nu}=0,\\g^{\mu\nu}t_{\mu\nu\rho}=0=g^{\mu\rho}t_{\mu\nu\rho}.
\end{align}
By contracting equation \eqref{eq15} over $\mu$ and $\rho$ we obtain $Q_\mu=T_\mu$.
Assuming that  $t_{\mu\nu\rho}\equiv 0$ \cite{WCW}, one may formulate the  Weitzenb\"{o}ck condition as
\begin{align}\label{eq17}
R=-6S_\mu S^\mu+\f{8}{3}T.
\end{align}
Now in equations \eqref{eq13} and \eqref{eq14}  the terms with coefficient $(1+\lambda)$ can  be simplified to
\begin{align}\label{eq18}
T_\mu^{~\rho\beta}+T^{\beta\rho}_{~~\mu}+T^{\rho~\beta}_{~\mu}-4T^{[\rho}\delta^{\beta]}_\mu=-\f{8}{3}T^{[\rho}\delta^{\beta]}_\mu-\epsilon_\mu^{~\rho\beta\sigma}S_\sigma,
\end{align}
and
\begin{align}\label{eq19}
2T^{\alpha\beta}_{~~~\nu}&T_{\alpha\beta\mu}-T_\mu^{~\alpha\beta}T_{\nu\alpha\beta}+2T_{\beta\alpha(\mu}T^{\alpha\beta}_{~~~\nu)}-4T_\mu T_\nu\nonumber\\
&=-\f{24}{9}T_\mu T_\nu+2(S_\alpha S^\alpha g_{\mu\nu}-S_\mu S_\nu),
\end{align}
respectively. Taking the trace of equation \eqref{eq13} over indices $\beta$ and $\mu$, we have
\begin{align}\label{eq20}
\alpha\nabla_\alpha T^{\rho\alpha}-\beta T^\rho\Box T=\f{4}{3}(1+\lambda)T^\rho-2 w^\rho.
\end{align}
Now, substituting the LHS of the above equation into \eqref{eq13} we obtain
\begin{align}\label{eq21}
(1+\lambda)\epsilon_\mu^{~\rho\beta\sigma}S_\sigma=0.
\end{align}

If one assumes $\lambda\neq -1$ then $S_\mu=0$.
We note that from equation \eqref{eq17} one has $R=8/3T$ which implies that the vector $T^\mu$ should be space-like for the accelerating Universe with $R=6(\dot{H}+2H^2)$, where $H$ is the Hubble parameter.

\subsection{The case $\alpha =0$}

In order to further simplify  the gravitational field equations of the WCW model with a Lagrange multiplier, let us assume that $\alpha=0$, as in \cite{WCW}. In this case from equation \eqref{eq20} we find
\begin{align}\label{eq22}
\Box T=-\f{4}{3\beta}(1+\lambda)+\f{2}{\beta T}w_\rho T^\rho,
\end{align}
provided that $T\neq 0$. Substituting (\ref{eq20}) into \eqref{eq13} we obtain
\begin{align}\label{eq23}
T^\alpha T_\alpha(w^\rho\delta^\beta_\mu-w^\beta\delta^\rho_\mu)=w^\alpha T_\alpha(T^\rho\delta^\beta_\mu-T^\beta\delta^\rho_\mu),
\end{align}
which implies that $T_\mu=Aw_\mu$, where $A$ is a constant. In order to obtain the value of the constant $A$, we take the covariant divergence of equation  \eqref{eq12}, with the result
\begin{align}\label{eq24}
\nabla_\mu(6w^\mu-4T^\mu)=0.
\end{align}
The above equation implies that $A=3/2$, so we conclude that
\begin{align}\label{eq25}
T_\mu=\f{3}{2}w_\mu.
\end{align}
Substituting the above equation into \eqref{eq12} we obtain the dynamical field equation of the Weyl vector
\begin{align}\label{eq26}
\Box w_\mu-\nabla_\mu\nabla_\nu w^\nu-w^\nu R_{\nu\mu}=0.
\end{align}
Now, using \eqref{eq25}, we write equation \eqref{eq13} as
\begin{align}\label{eq27}
\Box T=-\f{4}{3\beta}\lambda,
\end{align}
which implies
\begin{align}\label{eq28}
\lambda=-\f{27\beta}{16}\Box w^2,
\end{align}
where $w^2=w_\alpha w^\alpha$. Substituting $T^\mu$ and $\lambda$ from equations \eqref{eq25} and \eqref{eq28} into the metric field equation we obtain the effective energy-momentum tensor of the WCW model with the Lagrange multiplier,   equation \eqref{eq14-2}, as
\begin{widetext}
\begin{align}\label{eq29}
T^{eff}_{\mu\nu}=\left(1-\f{27\beta}{16}\Box w^2\right)^{-1}&\bigg[\f{\kappa^2}{2}T^m_{\mu\nu}-\f{27\beta}{16}(\nabla_\mu\nabla_\nu\Box w^2-\Box^2 w^2 g_{\mu\nu})+\f{\kappa^2}{2}\left(W_{\mu\alpha}W_\nu^{~\alpha}-\f{1}{4}W_{\alpha\beta}W^{\alpha\beta}g_{\mu\nu}\right)\nonumber\\
&+\f{81\beta}{32}(2w^2\Box w^2 g_{\mu\nu}-2\nabla_\mu w^2\nabla_\nu w^2+g_{\mu\nu}\nabla_\alpha w^2\nabla^\alpha w^2)\bigg].
\end{align}
\end{widetext}

In summary, one may obtain the Weyl vector from equation \eqref{eq26} and then the Lagrange multiplier $\lambda$ from equation \eqref{eq28}. The field equation \eqref{eq14-1}, together with equation \eqref{eq29} can then be used to obtain the evolution of the metric. Hence a complete solution of the gravitational field equations in the WCW model with a Lagrange multiplier can be constructed, once the thermodynamic parameters of the matter (energy density and pressure) are known.

It is worth mentioning that because of the general covariance, the matter energy-momentum tensor should be conserved due to the Bianchi identity. One can easily prove this statement in the case $\alpha=0$. Using equation \eqref{eq29}, one may write equation \eqref{eq14-1} as
\begin{widetext}
\begin{align}\label{eq29-1}
\left(1-\f{27\beta}{16}\Box w^2\right)G_{\mu\nu}=&\bigg[\f{\kappa^2}{2}T^m_{\mu\nu}-\f{27\beta}{16}(\nabla_\mu\nabla_\nu\Box w^2-\Box^2 w^2 g_{\mu\nu})+\f{\kappa^2}{2}\left(W_{\mu\alpha}W_\nu^{~\alpha}-\f{1}{4}W_{\alpha\beta}W^{\alpha\beta}g_{\mu\nu}\right)\nonumber\\
&+\f{81\beta}{32}(2w^2\Box w^2 g_{\mu\nu}-2\nabla_\mu w^2\nabla_\nu w^2+g_{\mu\nu}\nabla_\alpha w^2\nabla^\alpha w^2)\bigg].
\end{align}
\end{widetext}
Taking the divergence of the above equation one  obtains
\begin{align}\label{eq29-2}
-\f{27\beta}{16}&\nabla^\mu\Box w^2 G_{\mu\nu}=\f{\kappa^2}{2}\nabla^\mu T^m_{\mu\nu}\nonumber\\&-\f{27\beta}{16}\big(\Box\nabla_\nu\Box w^2-\nabla_\nu\Box^2 w^2\big)+\f{81\beta}{16}w^2\nabla_\nu\Box w^2.
\end{align}
Now, using the identity
\begin{align}\label{eq29-3}
\nabla^\mu\nabla_\nu A_\mu-\nabla_\nu\nabla^\mu A_\mu=R^\alpha_\nu A_\alpha,
\end{align}
and considering the Weitzenb\"ock condition which reads $R=6w^2$, where $R$ is the Ricci scalar, one  easily finds $\nabla^\mu T^m_{\mu\nu}=0$.

\section{The limiting case $w^\mu=0$ and the teleparallel gravity}\label{w}

In this Section we consider the limiting case in which the Weyl vector becomes zero. We also assume $\alpha=0$ for simplicity. The action of the theory becomes
\begin{align}\label{w1}
S&=\f{1}{\kappa^2}\int d^4x\sqrt{-g}\bigg[\beta\nabla_\mu T\nabla^\mu T+\kappa^2L_m\nonumber\\&+(1+\lambda)\big(R+T^{\mu\alpha\nu}T_{\mu\alpha\nu}+2T^{\mu\alpha\nu}T_{\nu\alpha\mu}-4T_\mu T^\mu\big)\bigg],
\end{align}
One may then obtain the field equations for the torsion tensor and the metric as
\begin{align}\label{w2}
(1+\lambda)\big(T_\mu^{~\rho\beta}+T^{\beta\rho}_{~~\mu}+T^{\rho~\beta}_{~\mu}&-4T^{[\rho}\delta^{\beta]}_\mu\big)\nonumber\\&-2\beta T^{[\rho}\delta^{\beta]}_\mu\Box T=0,
\end{align}
and
\begin{align}\label{w3}
G_{\mu\nu}=T^{eff}_{\mu\nu},
\end{align}
with
\begin{widetext}
\begin{align}\label{w4}
T^{eff}_{\mu\nu}&=(1+\lambda)^{-1}\bigg[ \f{\kappa^2}{2}T^m_{\mu\nu}+\nabla_\mu\nabla_\nu\lambda-g_{\mu\nu}\Box\lambda-(1+\lambda)\bigg(2T^{\alpha\beta}_{~~~\nu}T_{\alpha\beta\mu}-T_\mu^{~\alpha\beta}T_{\nu\alpha\beta}+2T_{\beta\alpha(\mu}T^{\alpha\beta}_{~~~\nu)}-4T_\mu T_\nu\bigg)\nonumber\\
&-\beta\left(\nabla_\mu T\nabla_\nu T-\f{1}{2}g_{\mu\nu}\nabla_\alpha T\nabla^\alpha T-2 T_\mu T_\nu\Box T\right)+\f{1}{2}(1+\lambda)\left(T^{\gamma\beta\alpha}T_{\gamma\beta\alpha}+2T^{\gamma\beta\alpha}T_{\alpha\beta\gamma}-4T^\alpha T_\alpha\right)g_{\mu\nu}\bigg],
\end{align}
\end{widetext}
The variation of the action with respect to the  Lagrange multiplier gives the Weitzenb\"{o}ck condition \eqref{eq10}. Now consider the decomposition of the torsion tensor, given by equation \eqref{eq15}, with $t_{\mu\nu\rho}=0$. One can again obtain $S_\mu=0$ by the same trick as in Section III. We then obtain the Weitzenb\"{o}ck condition in the form
\begin{align}\label{w5}
R=\f{8}{3}T.
\end{align}
From equation \eqref{w2} one can isolate the Lagrange multiplier
\begin{align}\label{w6}
\lambda=-\f{3\beta}{4}\Box T-1.
\end{align}
By substituting equation \eqref{w6} one can check that equation \eqref{w2} is automatically satisfied. The metric then equation becomes
\begin{align}\label{w7}
\Box T G_{\mu\nu}&=-\f{2\kappa^2}{3\beta}T^m_{\mu\nu}+\nabla_\mu\nabla_\nu\Box T-g_{\mu\nu}\Box^2 T\nonumber\\&+\f{4}{3}\nabla_\mu T\nabla_\nu T-\f{2}{3}g_{\mu\nu}\nabla_\alpha T\nabla^\alpha T-\f{4}{3}g_{\mu\nu}T\Box T.
\end{align}

\subsection{The case $\beta=0$}

For $\beta =0$ the torsion has no kinetic term. Putting $\beta=0$ in equation \eqref{w2} and using equation \eqref{eq18}, we obtain $T^\rho=0$. The trace of equation \eqref{w2} then gives $S_\mu=0$. Therefore, from the field equations  we obtain $T^\mu_{~\rho\nu}=0$  and the theory reduces to a Brans-Dicke type theory, with equations of motion
\begin{align}\label{w8}
G_{\mu\nu}=(1+\lambda)^{-1}\bigg[\f{\kappa^2}{2}T^m_{\mu\nu}+\nabla_\mu\nabla_\nu\lambda-g_{\mu\nu}\Box\lambda\bigg],
\end{align}
and
\begin{align}\label{w9}
\Box\lambda=\f{\kappa^2}{6}T^m,
\end{align}
respectively, where $T^m$ is the trace of the energy-momentum tensor. We have used the Weitzenb\"{o}ck condition $R=0$ to obtain equation \eqref{w9}.

\section{The Newtonian Limit and the generalized Poisson equation}\label{weak}

In this Section, we will obtain the generalized Poisson equation describing the weak field limit of the WCW theory with Lagrange multiplier. Taking the trace of equation \eqref{eq14}, using the  Weitzenb\"{o}ck condition \eqref{eq17}, and noting that $S_\mu=0$ in our setup, we obtain
\begin{align}
\f{1}{2}\kappa^2 T^m&-3\Box\lambda-6w^2+8T^\mu w_\mu\nonumber\\&+\beta(\nabla_\alpha T\nabla^\alpha T+2T\Box T)=0,
\end{align}
Now, using equation \eqref{eq20} to eliminate the $\Box T$ term, we find
\begin{align}
(1+\lambda)R&=\f{1}{2}\kappa^2 T^m-3\Box\lambda-6w^2+\beta\nabla_\mu T\nabla^\mu T\nonumber\\&+12w_\mu T^\mu+2\alpha T_\mu\nabla_\nu T^{\mu\nu}.
\end{align}
In the limit of the weak gravitational fields  the $(00)$ component of the metric tensor takes the form $g_{00}=-(1+2\phi)$, where $\phi$ is the Newtonian potential. In this limit we have $R=-\nabla^2\phi$, and obtain the generalized Poisson equation as
\begin{align}
\nabla^2\phi=(1+\lambda)^{-1}\bigg[\f{1}{4}\kappa^2\rho&+\f{3}{2}\Box\lambda+3w^2-6w_\mu T^\mu\nonumber\\&-\alpha T_\mu\nabla_\nu T^{\mu\nu}\bigg].
\end{align}
In obtaining the above equation we have assumed that the matter content of the Universe is pressureless dust, and we have used the  Weitzenb\"{o}ck equation to keep terms up to first order in $\phi$.

In the particular case $\alpha=0$, from equation \eqref{eq28} we find  that the Lagrange multiplier is of the order of $\phi$. Using equation \eqref{eq25} we obtain the generalized Poisson equation as
\begin{align}
\nabla^2\phi=\f{1}{4}\kappa^2\rho-\f{81}{32}\beta\Box^2 w^2+6w^2.
\end{align}

For $w=0$, we recover the standard Poisson equation of Newtonian gravity.
                              \section{Cosmological Solutions}\label{cos}
In this Section we consider the cosmological solutions and implications of the WCW model with Lagrange multiplier. We assume  that the metric of the space-time has the form of the flat Friedmann-Robertson-Walker (FRW) metric,
\begin{align}\label{cos1}
 ds^2=-dt^2+a^2(t)\left(dx^2+dy^2+dz^2\right).
\end{align}
Also in the following we suppose that the tensor $t_{\mu \nu \rho}$ vanishes, $t_{\mu\nu\rho}=0$. As we have mentioned in the previous Section, $S_\mu=0$ and $T_\alpha$ should be space-like in order to obtain an accelerating solution. We consider only models in which the Universe is filled with a perfect fluid, with the energy-momentum tensor given in a  comoving frame by
\begin{align}\label{cos6}
 T^\mu_\nu=\textmd{diag}(-\rho,p,p,p),
\end{align}
where $\rho $ and $p$ are the thermodynamic energy density and pressure, respectively. The Hubble parameter is defined as $H=\dot{a}/a$. As an indicator of the accelerated expansion we will consider the deceleration parameter $q$, defined as
\be
q=\frac{d}{dt}\frac{1}{H}-1.
\ee
If $q<0$, the Universes experiences an accelerated expansion while $q>0$ corresponds to a decelerating dynamics.
\subsection{The case $\alpha=0$}
In this case the cosmological dynamics is described by  equation \eqref{eq26} which represents the dynamical equation for the Weyl vector together with equations \eqref{eq28} and \eqref{eq14-1} which determine the Lagrange multiplier and the scale factor, respectively. The  Weitzenb\"{o}ck condition is
\begin{align}\label{cos1-1}
R=6w^2.
\end{align}
Let us assume that the Weyl vector is of the form
\begin{align}\label{cos1-2}
w_\mu&=a(t)\psi(t)(0,1,1,1).
\end{align}
The  Weitzenb\"{o}ck equation reduces to
\begin{align}\label{cos1-3}
\dot H+2H^2-3\psi^2=0,
\end{align}
and the Lagrange multiplier can be obtained as
\begin{align}\label{cos1-4}
\lambda=\f{81\beta}{8}\big(2\Psi^2+\dot\Psi+3\Psi H\big)\psi^2,
\end{align}
where we have defined $\Psi=\dot\psi/\psi$.
The dynamical equation for the Weyl vector is
\begin{align}\label{cos1-5}
\dot\Psi+\dot H+\Psi^2+2H^2+3\Psi H=0.
\end{align}
The off diagonal elements of the metric field equation gives
\begin{align}\label{cos1-6}
\Psi+H=0.
\end{align}
One can then check that the Weyl equation \eqref{cos1-5} is automatically satisfied. By substituting $H$ from \eqref{cos1-6} to the diagonal metric equations one obtains
\begin{align}\label{cos1-7}
\f{3}{8}\beta\Psi\ddot\Psi+\f{3}{8}\beta(3\psi^2&-\Psi^2)\dot\Psi-\f{3}{8}\beta(3\psi^2-\Psi^2)\Psi^2\nonumber\\
&-\f{1}{27}\psi^{-2}\Psi^2+\f{1}{162}\kappa^2\psi^{-2}\rho=0,
\end{align}
and
\begin{align}\label{cos1-8}
\f{1}{8}\beta \dddot{\Psi}&-\f{1}{8}\beta(2\dot\Psi+9\psi^2+\Psi^2)\dot\Psi-\f{2}{81}\psi^{-2}\dot\Psi\nonumber\\&-\f{3}{8}\beta\Psi^4+\f{1}{27}\psi^{-2}\Psi^2+\f{1}{162}\kappa^2\psi^{-2}p=0,
\end{align}
We note that in this case we have four equations, \eqref{cos1-3}, \eqref{cos1-6}, \eqref{cos1-7} and \eqref{cos1-8} for four unknowns $a$, $\psi$, $\rho$ and $p$. The Lagrange multiplier can then be obtain from equation \eqref{cos1-4}. Equation (\ref{cos1-6}) can be immediately integrated to give
\be
a(t)\psi (t)={\rm constant}=C_0\neq 0,
\ee
where $C_0$ is an arbitrary constant of integration.
With the use of $\psi (t)=C_0/a(t)$, the  Weitzenb\"{o}ck condition, equation (\ref{cos1-3}), becomes
\be
a\ddot{a}+\dot{a}^2-3C_0^2=0,
\ee
or equivalently
\be
\frac{d}{dt}\left(a\dot{a}\right)=3C_0^2,
\ee
which immediately leads to
\be
a^2(t)=3C_0^2t^2+C_1t+C_2,
\ee
where $C_1$ and $C_2$ are arbitrary constants of integration.
By assuming the initial conditions $a(0)=a_0$ and $H(0)=H_0$, respectively, we obtain $C_2=a_0^2$, and $C_1=2a_0^2H_0$. Thus for the Hubble parameter we obtain
\be
H(t)=\frac{a_0^2 H_0+6 C_0^2 t}{a_0^2+2 a_0^2 H_0 t+6 C_0^2 t^2}.
\ee

The energy density of the Universe can be obtained from equation (\ref{cos1-8}) as
\begin{widetext}
\begin{align}
\kappa ^2\rho (t)&=\frac{4374 C_0^{12} t^{10}}{\left(2 a_0^2 H_0 t+a_0^2+3 C_0^2 t^2\right)^6}+\frac{2916 a_0^2 C_0^{10} t^8 (5 H_0 t+2)}{\left(2
   a_0^2 H_0 t+a_0^2+3 C_0^2 t^2\right)^6}+\frac{6 a_0^{12} H_0^2 (2 H_0 t+1)^4}{\left(2 a_0^2 H_0 t+a_0^2+3
   C_0^2 t^2\right)^6}+\nonumber\\
   &\frac{36 a_0^{10} C_0^2 H_0 t (2 H_0 t+1)^3 (4 H_0 t+1)}{\left(2 a_0^2 H_0 t+a_0^2+3 C_0^2
   t^2\right)^6}+\frac{54 a_0^8 C_0^4 t^2 (2 H_0 t+1)^2 \left(26 H_0^2 t^2+12 H_0 t+1\right)}{\left(2 a_0^2 H_0 t+a_0^2+3
   C_0^2 t^2\right)^6}+\nonumber\\
   &\frac{648 a_0^6 C_0^6 t^4 (2 H_0 t+1) \left(11 H_0^2 t^2+7 H_0 t+1\right)}{\left(2 a_0^2 H_0
   t+a_0^2+3 C_0^2 t^2\right)^6}+\frac{486 a_0^4 C_0^8 t^6 \left(41 H_0^2 t^2+32 H_0 t+6\right)}{\left(2 a_0^2 H_0
   t+a_0^2+3 C_0^2 t^2\right)^6}+\nonumber\\
   &\frac{243\beta }{4 \left(2 a_0^2 H_0 t+a_0^2+3 C_0^2 t^2\right)^6}\times\Bigg\{2430 a_0^2 C_0^{10} H_0 t^4+324 a_0^2 C_0^8 t^3 \left(8 a_0^2 H_0^2-9 C_0^2\right)+\nonumber\\
   &3 a_0^4 C_0^4 t^2
   \Bigg[a_0^4 H_0^4+6 a_0^2 C_0^2 H_0^2 (84 H_0-1)+C_0^4 (9-972 H_0)\Bigg]+a_0^6 C_0^2 \Bigg[a_0^4
   H_0^4 (48 H_0+1)-\nonumber\\
   &6 a_0^2 C_0^2 H_0^2 (24 H_0+1)+9 C_0^4 (6 H_0+1)\Bigg]+2 a_0^4 C_0^2 t \Bigg[a_0^6
   H_0^5+6 a_0^4 C_0^2 H_0^3 (36 H_0-1)+\nonumber\\
   &9 a_0^2 C_0^4 (1-60 H_0) H_0+81 C_0^6\Bigg]+1458 C_0^{12} t^5\Bigg\}.
\end{align}
The thermodynamic pressure is found in the form
\begin{align}
\kappa ^2p&=\frac{2 \left[a_0^4 H_0^2-6 a_0^2 C_0^2 H_0 t-6 a_0^2 C_0^2-9 C_0^4 t^2\right]}{\left(2 a_0^2 H_0
   t+a_0^2+3 C_0^2 t^2\right)^2}+\frac{81 C_0^2\beta }{4 \left(2 a_0^2 H_0 t+a_0^2+3 C_0^2 t^2\right)^5}\times \nonumber\\
   &\Bigg[-35 a_0^8 H_0^4+135 a_0^6 C_0^2 H_0^2-63 a_0^4 C_0^4+324 a_0^2 C_0^6 H_0 t^3+t \left(558 a_0^4 C_0^4
   H_0-150 a_0^6 C_0^2 H_0^3\right)+\nonumber\\
   &t^2 \left(837 a_0^2 C_0^6-117 a_0^4 C_0^4 H_0^2\right)+243 C_0^8 t^4\Bigg].
\end{align}
For $t=0$ we obtain the initial values of the density and pressure as
\be\label{r0}
\rho (0)=\rho _0=6 H_0^2+\frac{243 \beta  C_0^2 \left(48 a_0^4 H_0^5+a_0^4 H_0^4-144 a_0^2 C_0^2 H_0^3-6 a_0^2 C_0^2 H_0^2+54
   C_0^4 H_0+9 C_0^4\right)}{4 a_0^6},
   \ee

   and
   \bea\label{p0}
  p(0)=p_0= \frac{2 \left(a_0^2 H_0^2-6 C_0^2\right)}{a_0^2}-
  \frac{81 \beta  C_0^2 \left(35 a_0^4 H_0^4-135 a_0^2 C_0^2 H_0^2+63
   C_0^4\right)}{4 a_0^6},
   \eea
    \end{widetext}
   respectively. Once the initial conditions $\left(a_0,H_0,\rho_0,p_0\right)$ are known, from Eqs.~(\ref{r0}) and (\ref{p0}) the values of the integration constants can be determined. The deceleration parameter can be obtained as
   \be
   q(t)=a_0^2\frac{a_0^2 H_0^2-3  C_0^2}{\left(a_0^2 H_0+3 C_0^2 t\right)^2}.
   \ee
   If the initial values of the scale factor and Hubble parameter satisfy the condition $a_0H_0<\sqrt{3}C_0$, $q<0$ for all times then the Universe is in an accelerated expansionary phase. If $a_0H_0=\sqrt{3}C_0$, $q(t)\equiv 0$ then the Universe is in a marginally inflating state. Finally, the Lagrange multiplier for this model can be obtained as
   \be
   \lambda (t)=\frac{81 a_0^2 \beta  C_0^2 \left(a_0^2 H_0^2-3 C_0^2\right)}{8 \left[a_0^2 (2 H_0 t+1)+3 C_0^2 t^2\right]^3}.
   \ee

\subsection{The case $\alpha\neq0$}
We assume that the Weyl vector is space-like, mimicking the proportionality of the torsion and the Weyl vector as in the case $\alpha=0$. Let us assume that
\begin{align}\label{cos2}
T_\mu&=a(t)\phi(t)(0,1,1,1),\nonumber\\
w^\mu&=\f{\psi(t)}{a(t)}(0,1,1,1).
\end{align}
By substituting these forms of the torsion and Weyl vector into equation \eqref{eq13} we obtain, after some algebra, the Lagrange multiplier
\begin{align}\label{eqlam}
 \lambda=&\f{3}{4}(6\beta\phi^2-\alpha)\dot{\Phi}-\f{3}{4}\alpha\dot{H}+\f{9}{2}\beta \phi^2\left(2\Phi+3H\right)\Phi\nonumber\\
 &-\f{3}{4}\alpha(\Phi^2+2H^2+3 H\Phi)+\f{3}{2}\f{\psi}{\phi}-1,
\end{align}
where we have defined
\begin{align}\label{cos4}
 \Phi=\f{\dot{\phi}}{\phi}.
\end{align}

By using equation \eqref{eqlam}, the field equation \eqref{eq12} becomes
\begin{align}\label{cos5}
 \ddot{\psi}+3H\dot{\psi}+(\dot{H}+2H^2+12\kappa^{-2})\psi-8\kappa^{-2}\phi=0.
\end{align}
The  Weitzenb\"{o}ck equation takes the form
\begin{align}\label{eqdh}
\dot{H}=-2H^2+\f{4}{3}\phi^2.
\end{align}

Substituting equations \eqref{eqlam} and \eqref{eqdh} into \eqref{eq14}, one obtains
\begin{widetext}
\begin{align}\label{cos7}
9H(6\beta\phi^2-\alpha)\ddot{\Phi}&+6\bigg[(\alpha-6\beta\phi^2)(2\phi^2-6H^2-3H\Phi)+36\beta\phi^2 H\Phi\bigg]\dot{\Phi}-3\dot{\psi}\left(\kappa^2\dot{\psi}+2\kappa^2 H\psi-6\f{H}{\phi} \right)\nonumber\\&+8\phi^4(2\alpha-9\beta\Phi^2)
+216\beta \phi^2 H \Phi^2(\Phi+2H)+6\phi^2\Phi(4\alpha\Phi-27\beta H^3)+24\psi\phi+9\alpha H^2\Phi (3H-\Phi)\nonumber\\&-3\psi^2(\kappa^2 H^2+12)+18H\f{\psi}{\phi}(H-\Phi)=2\kappa^2\rho,
\end{align}
\begin{align}\label{cos8}
 -&9(\alpha-6\beta\phi^2)\phi\dddot\Phi-9\phi\big[\alpha(5H+2\Phi)-6\beta\phi^2(5H+8\Phi)\big]\ddot\Phi+18\ddot\psi+18(18\beta\phi^2-\alpha)\phi\dot\Phi^2\nonumber\\
 &+9\big[40\beta\phi^5+6(24\beta\Phi^2+30\beta H\Phi-7\beta H^2-2\alpha)\phi^3+\alpha H(7H-4\Phi)\phi-2\psi\big]\dot\Phi+3 \kappa^2 \phi\dot\psi^2\nonumber\\&+8(117\beta\Phi^2-81\beta H\Phi-2\alpha)\phi^5+18\big[12\beta(2\Phi+5H)\Phi^3-2(4\alpha+21\beta H^2)\Phi^2+(27\beta  H^2-2\alpha)H\Phi\big]\phi^3
 \nonumber\\&+72\psi\phi^2+\big[6 \kappa^2 p+9\alpha H^2\Phi^2+3\kappa^2 H^2\psi^2-81\alpha H^3\Phi-36\psi^2\big]\phi\nonumber\\&+6(6H-6\Phi+\kappa^2\phi H\psi)\dot\psi+18\psi\Phi^2-36\psi H\Phi-18\psi H^2=0,
\end{align}
and
\begin{align}\label{cos9}
\kappa^2\dot{\psi}(\dot{\psi}+2\psi H)+\psi^2 (\kappa^2H^2-12)+4\alpha \phi^2\left(\dot{\Phi}-H^2+H\Phi+\f{4}{3}\phi^2\right)+8\psi\phi=0,
\end{align}
\end{widetext}
Eqs.~\eqref{cos5}, \eqref{eqdh}, \eqref{cos7}, \eqref{cos8} and \eqref{cos9} form a closed system of differential equations for five unknowns $\psi$, $\phi$, $H$, $p$ and $\rho$. Equation \eqref{eqlam} can then be used to determine the Lagrange multiplier.

In the following we will look only for a de Sitter type solution of the field equations \eqref{cos5}-\eqref{cos9} with $H=H_0={\rm constant}$ and $a=\exp\left(H_0t\right)$, respectively. Then the  Weitzenb\"{o}ck condition  (\ref{eqdh}) immediately gives
\be
\phi ^2=\frac{3}{2}H_0^2={\rm constant},
\ee
and $\Phi \equiv 0$, respectively. Equation (\ref{cos5}) takes the form
\be
\ddot{\psi }+3H_0\dot{\psi}+\left(2H_0^2-12\kappa ^{-2}\right)\psi =-4\sqrt{6}\kappa ^{-2}H_0^2,
\ee
with the general solution given by
\bea
\psi (t)&=&\frac{2\sqrt{6} H_0^2}{6-H_0^2 \kappa ^2}+c_1 e^{ \frac{ -\kappa  \sqrt{H_0^2 \kappa ^2+48}-3 H_0 \kappa ^2}{2 \kappa ^2}t}+\nonumber\\
&&c_2 e^{\frac{\kappa  \sqrt{H_0^2 \kappa
   ^2+48}-3 H_0 \kappa ^2}{2 \kappa ^2}t},
   \eea
   where $c_1$ and $c_2$ are arbitrary constants of integration. The simplest case corresponds to the choice $c_1=0$, $c_2=0$, giving
   \be
   \psi =\frac{2\sqrt{6} H_0^2}{6-H_0^2 \kappa ^2}={\rm constant}.
   \ee
   By substituting this form of $\psi $ into equation (\ref{cos9}) we obtain the value of $\alpha $ as
   \be
   \alpha =\frac{12\kappa ^2}{\left(6-H_0^2\kappa ^2\right)^2}.
   \ee
   For the energy density of the Universe we obtain
   \bea
   \kappa ^2\rho   =\frac{72H_0^2(2\kappa^2H_0^2+3)}{\left(6-H_0^2 \kappa ^2\right)^2},
   \eea
\begin{align}
 \kappa ^2p   =\frac{72H_0^2(2\kappa^2H_0^2-3)}{\left(6-H_0^2 \kappa ^2\right)^2}.
\end{align}
One can see that the energy density and the pressure is positive if $H_0\geq 1/\kappa^2\sqrt{3/2}$.
  \subsection{Cosmological models with  $w^\mu=0$}

Finally,  we consider the cosmological implications of the WCW model with Lagrange multiplier with $w^\mu=0$. Assuming the following form for the torsion,
\begin{align}
T_\mu=a(t)\phi(t)\big[0,1,1,1],
\end{align}
the  Weitzenb\"{o}ck condition is formulated as
\begin{align}
R-8\phi^2=0.
\end{align}
The Lagrange multiplier can be obtained in the form
\begin{align}
\lambda+1=\f{9}{2}\beta\phi^2\big(\dot\Phi+2\Phi^2+3H\Phi\big),
\end{align}
where we have defined $\Phi=\dot\phi/\phi$. The metric field equations take the form
\begin{align}
\ddot\Phi&+2\dot\Phi(2H+3\Phi)-\f{4}{3}\f{\phi^2}{H}(\dot\Phi+\Phi^2+3H\Phi)\nonumber\\&+\Phi(4\Phi^2+3H^2+8H\Phi+3\dot H)-\f{\kappa^2}{27\beta}\f{\rho}{H\Phi^2}=0,
\end{align}
and
\begin{align}
 &\dddot\Phi+\ddot\Phi(5H+8\Phi)+3\Phi\ddot H-4\phi^2(\dot\Phi+3\Phi^2+3H\Phi)+8\Phi^4\nonumber\\&+4\dot H(2\dot\Phi+4\Phi^2+3H\Phi)+9H^2\Phi(H+2\Phi)+20H\Phi^3\nonumber\\&+3\dot\Phi(10H\Phi+8\Phi^2+3H^2+2\dot\Phi)+\f{\kappa^2}{9\beta}\phi^{-2}p=0,
 \end{align}
 respectively.

 Let us consider the case $a(t)=t^s$. In this case one  obtains
 \begin{align}\label{}
\phi(t)=\f{\sqrt{3s(2s-1)}}{t},
\end{align}
 and the energy density and  pressure take the form
 \begin{align}\label{}
\rho(t)&=\f{81s^2(3s^2+8s-10)(2s-1)}{4\kappa^2}\f{\beta}{t^6},\\
p(t)&=-\f{81(3s^3+2s^2-26s+20)(2s-1)s}{4\kappa^2}\f{\beta}{t^6}.
\end{align}
In order to have a consistent solution, $\phi$ should be real and $\rho$ and $p$ must be positive. This restricts the range of $s$ to
\begin{align}\label{}
\f{1}{3}(\sqrt{46}-4)<s<2.
\end{align}

For the deceleration parameter we obtain
\be
q=\frac{1}{s}-1,\quad -\frac{1}{2}<q<0.078.
\ee

 In the case $a(t)=e^{H_0t}$ we have
 \begin{align}\label{}
\phi(t)^2=\f{3}{2}H_0^2,
\end{align}
 with the matter energy density and the pressure becoming exactly zero
 \be
 \rho =p=0.
 \ee

                              \section{Conclusion}\label{conclu}
In this paper we have considered an extension of the  Weitzenb\"{o}ck type gravity models  formulated in a Weyl-Cartan space time. The basic difference between the present and the previous investigations is the way in which the  Weitzenb\"{o}ck condition which in a Riemann-Cartan space time  requires the exact cancellation of the Ricci scalar and the torsion scalar, is implemented. By starting with a general geometric framework, corresponding to a $CW_4$ space - time described by a metric tensor, torsion tensor and Weyl vector, we formulated the action of the gravitational field by including the  Weitzenb\"{o}ck condition via a scalar Lagrange multiplier. With the use of this action the gravitational field equations have been explicitly obtained. They show the explicit presence in the field equations of a new degree of freedom, represented by the Lagrange multiplier $\lambda $. The field equations must be consistently solved together with the  Weitzenb\"{o}ck condition which allows the unique
determination of the Lagrange multiplier $\lambda $. The weak field limit of the model was also investigated and it was shown that the Newtonian approximation leads to a generalization of the Poisson equation where besides the matter energy-density, the weak field gravitational potential also explicitly depends on the Lagrange multiplier and the square of the Weyl vector.

An interesting particular case is represented by the zero Weyl vector case. For this choice of the geometry the covariant divergence of the metric tensor is zero and the  Weitzenb\"{o}ck condition takes the form of a proportionality relation between the Ricci scalar and the torsion scalar, respectively. When one neglects the kinetic term associated to the torsion, the model reduces to a Brans-Dicke type theory where the role of the scalar field is played by the Lagrange multiplier.

The cosmological implications of the  theory have also been investigated by considering a flat FRW background type cosmological metric. We have considered three particular models, corresponding to the zero and non-zero values of the coupling constant $\alpha $, and to the zero Weyl vector respectively. For $\alpha =0$ the field equations can be solved exactly, leading to a scale factor of the form $a(t)=\sqrt{3c_0^2t^2+2H_0a_0^2+a_0^2}$. The energy density and the pressure are monotonically decreasing functions of time and are both non-singular at the beginning of the cosmological evolution. The nature of the cosmological expansion - acceleration or deceleration - is determined by the values of the constants $\left(C_0, a_0,H_0\right)$ and three regimes are possible: accelerating, decelerating, or marginally inflating. In the case $\alpha \neq 0$, we have considered only a de Sitter type solution of the field equations. Such a solution does exist if the matter energy density and pressure
are constants, or, more exactly, the decrease in the matter energy density and pressure due to the expansion of the Universe is exactly compensated by the variation in the energy and pressure due to the geometric terms in the energy-momentum tensor.

In the case of the cosmological models with vanishing Weyl vector we have investigated two particular models corresponding to a power law and exponential expansion, respectively. In the case of the power law expansion, the energy density and pressure satisfy a barotropic equation of state, so that $p\sim \rho$ where both the energy and pressure decay as $t^{-6}$. Depending on the value of the parameter $s$, both decelerating and accelerating models can be obtained. On the other hand, for a vanishing Weyl vector, the de Sitter type solutions require a vanishing matter energy density and pressure and hence the accelerated expansion of the  Universe is determined by the geometric terms associated with torsion which play the role of an effective cosmological constant.

In the present paper we have introduced a theoretical model for gravity, defined in a Weyl-Cartan space-time, in which the Weitzenb\"{o}ck geometric condition has been included in the action via a Lagrange multiplier method. The field equations of the model have been derived by using variational methods, and some cosmological implications of the model have been explored. Further astrophysical and cosmological implications of this theory will be considered elsewhere.


\appendix
\section{Note on Weyl gauge invariance}
Suppose that length of a vector at point $x$ is $l$. In the Weyl geometry, the length of the vector under parallel transportation to the nearby point $x^\prime$ is $l^\prime=\xi l$. On the other hand, the change in the length of the vector can be written as
\begin{align}\label{ap1}
\delta l=lw_\mu\delta x_\mu.
\end{align}
So, the change in the Weyl vector is
\begin{align}\label{ap2}
w_\mu\rightarrow w^\prime_\mu=w_\mu+\partial_\mu\log\xi,
\end{align}
From the above relations, one  obtains the change in the metric tensor
\begin{align}\label{ap3}
g_{\mu\nu}&\rightarrow g^\prime_{\mu\nu}=\xi^2 g_{\mu\nu},\\
g^{\mu\nu}&\rightarrow g^{\prime\mu\nu}=\xi^{-2} g^{\mu\nu}.
\end{align}
The torsion tensor is invariant under the above gauge transformation, i.e.,
\begin{align}\label{ap4}
T^\mu_{~\rho\sigma}\rightarrow T^{\prime\mu}_{~\rho\sigma}=T^\mu_{~\rho\sigma}
\end{align}
We note that the curvature tensor \eqref{eq6} is covariant with the power $-2$, which means
\begin{align}\label{ap5}
K^\prime=\xi^{-2}K.
\end{align}
and the metric determinant has power $4$. Naturally, one demands to make the Lagrangian \eqref{eq7} gauge-invariant. In order to do so one can add a scalar field $\beta$ or a Dirac field  with power $-1$ and write the first term in equation \eqref{eq7} as $\sqrt{-g}\beta^2 K$ to make it gauge-invariant. However, the  Weitzenb\"{o}ck condition \eqref{eq10} is neither gauge invariant nor covariant. In fact, one may write
\begin{align}\label{ap6}
\mathcal{W}^\prime=\xi^{-2}\mathcal{W}-6\left(\nabla_\nu k^\nu+k^\nu k_\nu\right),
\end{align}
where $\nabla$ is the metric covariant derivative and we have defined $k_\alpha=\partial_\alpha\log\xi$. In order to make the  Weitzenb\"{o}ck condition gauge-covariant, one should add to $\mathcal{W}$ some terms containing the torsion tensor and the Weyl vector. This generalization of the  Weitzenb\"{o}ck condition by adding torsion and Weyl tensors will be considered in our future work \cite{new}.

\end{document}